# Proximal binaural sound can induce subjective frisson

**Shiori Honda*[1], Yuri Ishikawa*[1], Rei Konno[1], Eiko Imai[2], Natsumi Nomiyama[2], Kazuki Sakurada[2], Takuya Koumura[3], Hirohito M. Kondo[4], Shigeto Furukawa[3], Shinya Fujii†[2], Masashi Nakatani†[2,5]**

[1] Graduate School of Media and Governance, Keio University, Fujisawa, Kanagawa, Japan

[2] Faculty of Environment and Information Studies, Keio University, Fujisawa, Kanagawa, Japan

[3] NTT Communication Science Laboratories, Atsugi, Kanagawa, Japan

[4] School of Psychology, Chukyo University, Nagoya, Aichi, Japan

[5] JST PRESTO

**\*** Both authors contributed equally to this study

**†** Corresponding authors

**Correspondence:**
Masashi Nakatani, Ph.D.
mn2598@sfc.keio.ac.jp

Shinya Fujii, Ph.D.

sfujii@sfc.keio.ac.jp

**Abstract**

Auditory frisson is the experience of feeling of cold or shivering related to sound in the absence of a physical cold stimulus. Multiple examples of frisson-inducing sounds have been reported, but the mechanism of auditory frisson remains elusive. Typical frisson-inducing sounds may contain a looming effect, in which a sound appears to approach the listener's peripersonal space. Previous studies on sound in peripersonal space have provided objective measurements of sound-inducing effects, but few have investigated the subjective experience of frisson-inducing sounds. Here we explored whether it is possible to produce subjective feelings of frisson by moving a noise sound (white noise, rolling beads noise or frictional noise produced by rubbing a plastic bag) stimulus around a listener's head. Our results demonstrated that sound-induced frisson can be experienced stronger when auditory stimuli are rotated around the head (binaural moving sounds) than the one without the rotation (monaural static sounds), regardless of the source of the noise sound. Pearson's correlation analysis showed that several

**Proximal binaural sounds make frisson**

acoustic features of auditory stimuli, such as variance of interaural level difference (ILD), loudness, and sharpness, were correlated with the magnitude of subjective frisson. We had also observed that the subjective feelings of frisson by moving a musical sound had increased comparing with a static musical sound.

**Introduction**

Sound frisson is an intriguing phenomenon which we experience in our daily life. For example, the buzz of a mosquito, perceived to be approaching to or moving around the body, can induce frisson. Similarly, a fly or bee buzzing in the same room as the listener, particularly around the listener's head, can have this effect. Thus, an approaching sound can be associated with frisson, regardless of the pitch of the sound (mosquitoes make a high-pitched noise and flies produce a low-pitched one). In this paper, we define frisson as the feeling of coldness or shivering in the absence of physically cold stimulus.

A sound that can induce subjective frisson and looming sounds may have in common. Looming sounds (sounds that seem to approach the hearer) are known to enhance human arousal (Bach et al., 2009). Changes in behavioral, physiological, or neurophysiological states caused by looming sounds are collectively referred to as the looming effect (Bach et al., 2009; Tajadura-Jimenez et al., 2010).

A recent study reported that approaching sounds can cause quicker responses than receding sounds in the simple reaction task. That study termed this the looming effect, describing it as the perception of approaching sounds as more salient than receding ones (Neuhoff, 2016). Moreover, ~~the human voice or~~ a pure tone or the tonal vowel stimulus elicited a stronger looming effect than white noise (~~Barratt and Davis, 2015~~ McCarthy and Olsen, 2017). It was found in the previous studies that the velocity of the looming effect is influenced by two factors: the positional relationship between the participant and auditory stimuli in 3D space, and the types of sound stimuli (e.g., tonal sound, white noise, human voice, and music).

The looming effect is also experienced in response to sounds that seem to approach in three-dimensional (3D) space. Bach et al. (2009) reported a spatial looming effect, in which participants exhibited skin-conductance responses, indicating that spatial perceptions of sound can evoke changes in phasic alertness and in the physiological state of the listener. Furthermore, binaural (dichotic or stereo) sounds can induce an auditory perception of motion (Altman, 1988) or the perception of sound approaching (Okada and Hirahara, 2015). To the best of our knowledge, there has been no work on the production of the looming effect by binaural sounds, but it is plausible that the looming effect can be elicited if the auditory stimulus is perceived to be approaching. These findings imply that the looming effect can be produced by both monaural (diotic; presenting the same acoustic stimulus in both ears) and binaural (dichotic; presenting different acoustic stimuli for each ear) auditory stimuli.

Multimedia content that produces autonomous sensory meridian response (ASMR) can also induce frisson-like sensations. ASMR has received close attention in online communities hosted on YouTube and Reddit, and several conceptions of its nature have been put forward (Barratt et al., 2017). In this paper, we define ASMR as the simultaneous perception of frisson and pleasantness.

The relationship between ASMR-inducing multimedia content and looming sounds appears clear: both auditory stimuli are to be in spatial motion and both can produce frisson. Barratt and Davis (2015) and Barratt et al. (2017) have reported that ASMR can be associated with sensory tingling or frisson. They have found that ASMR can be produced by a range of sounds, including whispering voices, crisp sounds (including the rustling of metallic foil and tapping fingernails), and the sounds produced by slow movements. ASMR-inducing multimedia content and looming sound both have the feature that the sound stimulus appears to be in spatial motion.

Frisson has also been discussed in the context of musical experiences (Sloboda, 1991), where it has been identified as an "affect that shows close links to musical surprise" (Huron and Margulis, 2011; see also Harrison and Loui, 2014). Recent neuroimaging studies have reported neural correlates of frisson while listening to music (Blood and Zatorre, 2001; Salimpoor et al., 2011). These studies further reported that music can change the subjective evaluation of frisson.

As noted above, music itself can produce frisson; however, the feeling of moving sound generated by audio mixing and/or listening to music in 3D space with loudspeakers may be able to enhance the experience. For instance, panning a sound source from a left speaker to a right speaker can give an impression of motion through sound (Carlile and Leung, 2016). Alternatively, witnessing a live performance can produce a stronger subjective frisson than listening to a recorded version of the same material, possibly because the performers are creating sound in front of an audience that is reflected in a 3D space. Furthermore, performers in such a context may move back and forth on the stage and thus generate multiple auditory cues that the audience can use to localize the source of the sound. The moving sound thus produced can facilitate frisson experiences.

Here, we explored the factors that influence subjective frisson, particularly in two areas: the positional relationship between the participant and the auditory stimuli in 3D space and the types of auditory stimuli (i.e., three kinds of non-musical noise sound and one musical sound). Then, we tested the hypothesis that proximal moving sound (e.g., moving around the participant's head) can induce subjective frisson. We also tested whether sounds with detailed positional information (e.g., binaural sounds that are recorded via a dummy head) could elicit stronger frisson than those without this information. Moreover, we explored the influence of personal traits such as anxiety and impulsive behaviors on sound induced frisson, because the previous study reported that behavioral responses to noise sounds may have individual differences (Riskind et al., 2014, Fredborg et al., 2017).

**Methods**

*Participants:*

The participants in this study were 19 healthy students (11 women and 8 men; mean age, 19.9 years; SD, 1.24 years; age range, 18–21 years). They had normal hearing without hearing loss, as confirmed by audiometry (Audiometer, cat# 7-4910-01, Navis). The experimental protocol was approved by the Ethical Committee on Human Experimentation of Keio University, Shonan Fujisawa Campus (approval # 183). Each participant provided written informed consent prior to participation in the experiment.

*Sound stimuli:*

We used three non-musical stimuli: white noise, rolling glass beads (the diameter of the hole in the center of each bead was 16.4 mm, and the diameter of bead as a whole was 63.9 mm), plastic bag (with a thickness of 0.01 mm), and one musical stimulus ("Strobe Hello," recorded by Powapowa-P). We compared the frisson rating between some musical and non-musical stimuli in our pilot study. The results of the pilot study showed that the frisson rating of the participants showed a gradation between musical and non-musical stimuli. In addition, there were large individual differences among musical stimuli. On the other hand, noise stimuli featured few individual differences, so we attempted to

**Proximal binaural sounds make frisson**

regulate the differences among musical stimuli using a song that had both music and non-music (noise) components.

The non-musical stimuli were recorded using a microphone (1/2 inch Free-field Microphone Type 4191 manufactured by Bruel & Kjaer Sound & Vibration) via an audio interface (Quad Capture manufactured by Roland). The sampling rate for both the musical and non-musical stimuli was 44.1 kHz. The musical stimulus was downloaded from piapro (https://piapro.jp/t/cT5y). To create binaural stimuli, the non-musical and musical sound stimuli were amplified (Audio Amplifier PM-5005/FN manufactured by Marantz), played through a loudspeaker (Alpair7v3 manufactured by Mark Audio), and recorded using a binaural microphone (Free Space Pro ll manufactured by 3DIO). The distance between the loudspeaker and the binaural microphone was 6 cm (Figure 1). This binaural microphone featured two separate microphones shaped like auricles and was lighter than a dummy-head recording system. This setup enabled rotation using a stepping motor, which we used to create the appearance sound stimuli rotating closely around the head, as shown in Figure 1. In this configuration, recorded stimuli were produced that sounded as if they were being played with multiple loudspeakers surrounding the head as in the previous study, but our configuration provided a smoother moving sound. We generated monaural sounds from binaural stimuli using the built-in Convert Stereo to Monaural function of Audacity (version 2.0.5), in which both the right and left sides of the sound were simultaneously presented to both ears. There were three rotational conditions for the recording: no rotation (0°/s), low speed (18°/s = 2 cm/s), and high speed (72°/s = 8 cm/s). Thus, there were 24 sound stimuli (four sound types [white noise, rolling beads, plastic bag, and music] × three rotation speeds [no rotation, low speed, and high speed] × two sound presentation conditions [monaural and binaural]). The length of each stimulus was 30 seconds.

The intensity of the sound stimuli was assessed using an artificial ear (TYPE2015, ACO Co., Ltd) via a Fast Fourier Transform analyzer (Type 6240, ACO Co., Ltd). We confirmed that A-weighted sound pressure levels of all the auditory stimuli were below 85 dB, a permissible noise level for 8 hours exposure (US Department of Health and Human Services, 1998) (Supplemental Table 1). In addition, we recorded sound stimuli for each ear via these devices to calculate the acoustic features.

------------------------------------------------- Figure 1 about here -------------------------------------------------

*Procedures:*

The participants were seated in a comfortable chair. Before the experiment began, the participants completed two questionnaires to assess their personalities: the State-Trait Anxiety Inventory-JYZ (Spielberger, 2010) and the UPPS-P Impulsive Behavior Inventory in Japanese (Whiteside and Lynam, 2001; Kondo et al., 2017). In the experiment, the sound stimuli were repeated twice in 48 trials. The order of stimuli was randomized across the participants. The 48 trials were divided into two blocks (24 trials each). A short break (approximately 5 min) was provided between the blocks. For each trial, a sound stimulus was played using headphones (QuietComfort 15 Acoustic Noise Cancelling Headphones, Bose Corporation) via a computer (HP ProBook 450 G3 Notebook PC, HP Inc.). The participants were instructed to fixate on a black cross mark drawn on a piece of paper while listening to the sound stimuli. After listening to each sound, they were instructed to report their subjective rating of frisson (from 0 [no frisson] to 1 [frisson]), using a visual analog scale (VAS). They were also instructed to report the level of pleasantness experienced (from −3 [unpleasant] to 3 [pleasant]) using the VAS. The next trial began after the completion of the subjective reports. The VAS was implemented with the MATLAB built-in GUI Development Environment and dialog box functions in MATLAB 2017b (MathWorks, Inc.). After 48 trials had been completed, the experimenter asked whether the binaural stimuli were perceived as spatial sounds rotating around the head or a stereo sound moving left and right. All participants reported perceiving the binaural sounds as spatial sounds. The participants also completed a questionnaire on demographic information (including age, sex, and years of musical experience) following the experiment.

*Analysis:*

We analyzed the participants' responses to non-musical and musical sounds separately, and only the first block of data was used because the participants tended to show a habituation effect after repetitive presentations of auditory stimuli. The z-score was used for the frisson and pleasantness ratings for each participant. For non-musical sounds, three within-participant factors were observed in the analysis of variance (ANOVA): three rotation speeds (no rotation, low speed, and high speed) two presentation conditions (monaural and binaural) × three sound types (white noise, rolling beads, and plastic bag). For the musical sound, two within-participant factors were observed in the ANOVA: three rotation speeds (no rotation, low speed, and high speed) × two presentation conditions (monaural and binaural). Pair-wise comparison was conducted as a post-hoc test. We performed Bonferroni correction to regulate the multiple comparisons.

We also calculated the acoustic features of the sound stimuli for each of left and right channels presented from headphones (i.e., non-stationary loudness [ISO532-1: 2017], sharpness, and tonality) using Oscope 2 (Ono Sokki Co., Ltd.)(Zwicker and Fastl, 2013). Then we calculated the arithmetic means of these acoustic features calculated from the left and the right channel, respectively. The interaural level difference (ILD) was calculated from original sound source files by using the ILD function (ILD values were computed with a Hanning window duration of 20 ms and a step size of 10 ms, passed through the 32-channel gammatone filterbank spaced between 1 and 22500 Hz) in the Two!Ears Auditory Front-End (https://github.com/TWOEARS/auditory-front-end) library run in MATLAB R2017b. The Pearson's correlation coefficient was calculated between these acoustic features of non-musical stimuli and their subjective ratings of frisson or pleasantness. Statistical analysis was conducted using IBM SPSS Statistics (Version 24).

**Results**

The mean of subjective ratings (frisson/pleasantness) of non-musical (noise) sounds and musical sounds is summarized in Table 1 and illustrated in Figure 2 ~~and Table 1~~, including the analysis of acoustic features. The result of correlation analysis between subjective frisson / pleasantness ratings and acoustic features are shown in Figure 3 and Supplemental Table 3. The details of the results are described in the following according to sound types (non-musical and musical sounds). We did not find any significant correlations between the personality trait measures and subjective frisson / pleasantness ratings (Supplementary Table 2).

*Non-musical sounds:*

The relationship between noise sounds stimuli and the frisson ratings is presented in Figure 2A. The three-way repeated ANOVA indicated no significant interaction among the three factors ($F(4,72) = 1.45$, $p = 0.23$, $\eta_p^2 = 0.075$) (see Table 2 for the details of statistical analysis). None of the interactions between the two factors were significant (sound type × presentation condition, $F(2,36) = 2.78$, $p = 0.075$, $\eta_p^2 = 0.13$; presentation condition × rotation speed, $F(2,36) = 2.76$, $p = 0.077$, $\eta_p^2 = 0.13$; rotation speed × sound type, $F(4,72) = 1.20$, $p = 0.32$, $\eta_p^2 = 0.062$). The main effect of the sound type was not significant ($F(2,36) = 2.20$, $p = 0.13$, $\eta_p^2 = 0.11$). The main effect of the presentation condition was

**Proximal binaural sounds make frisson**

significant (F(1,18) = 10.28, p < .01, $\eta_p^2$ = 0.36). The main effect of rotation speed was also significant (F(2,36) = 43.38, p < .001, $\eta_p^2$ = 0.71). The post-hoc analysis showed a significant difference between no rotation and rotation (static vs low-speed: p < .001, static vs high-speed: p < .001) but not between low-speed and high-speed rotation (p = 0.35).

-------------------------------------------------- Figure 2 about here --------------------------------------------------

The relationship between noise sounds stimuli and the pleasantness ratings is shown in Figure 2B. The three-way repeated ANOVA found no significant interaction among the three factors (F(4,72) = 0.27, p = 0.90, $\eta_p^2$ = 0.015). None of interactions between the two factors were significant (sound type × presentation condition, F(2,36) = 1.38, p = 0.26, $\eta_p^2$ = 0.071; rotation speed × presentation condition, F(2,36) = 0.51, p = 0.61, $\eta_p^2$ = 0.027; sound type × rotation speed, F(4, 72) = 0.55, p = 0.70, $\eta_p^2$ = 0.029;). A significant main effect was found for sound type (F(2,36) = 3.45, p < .05, $\eta_p^2$ = 0.16). The post-hoc analysis showed a significant difference between the plastic bag and white noise sounds (p < .05) but not between beads and plastic bag or beads and white noise sounds (p = 0.98 and p = 0.50, respectively). The main effect of presentation condition was not significant (F(1,18) = 2.29, p = 0.15, $\eta_p^2$ = 0.11). The main effect of rotation speed was significant (F(2,36) = 7.73, p < .01, $\eta_p^2$ = 0.30). The post-hoc analysis showed a significant difference between the no-rotation and with-rotation sounds (static vs low-speed: p < .01, static vs high-speed: p < .05), but not between low-speed and high-speed rotation (p = 1.00, respectively). We have also evaluated the effect of the sound pressure level of the auditory stimuli (-12 dB and + 2 dB) on subjective frisson and pleasantness ratings, but we did not find any statistically significant difference (Supplemental Figure 2)

The summary of the acoustic parameters is shown in Table 1. Correlation analysis indicated that the variances of ILD of the sound stimuli at 128 and 1336 Hz were correlated with the subjective magnitude of frisson (r = 0.93, p < .001 and r = 0.91, p < .01) and pleasantness (r = -0.76, p < .05 and r = -0.74, p < .05) (see Figure 3 and Supplemental Table 3). The loudness was negatively correlated with the subjective magnitude of frisson in the binaural sounds (r = -0.78, p < .05). The sharpness was positively correlated with the magnitude of frisson in both monaural and binaural sounds (r = 0.77, p < .05 and r = 0.95, p < .001, respectively) and negatively correlated with pleasantness in the binaural sounds (r = -0.76, p < .05). The tonality parameter did not significantly correlate with either the magnitude of frisson or pleasantness.

*Musical sounds:*

Figure 2C and 2D show the relationship between musical stimuli and the subjective magnitude of frisson and pleasantness. A two-way ANOVA showed significant interaction between the presentation condition and rotation speed for frisson rating (F(2,36) = 6.39, p < .01, $\eta_p^2$ = 0.26) (Figure 2C). The main effect of the presentation condition (F(1,18) = 10.77, p < .01, $\eta_p^2$ = 0.37) and rotation speed (F(2,36) = 11.45, p < .001, $\eta_p^2$ = 0.39) was significant. A post-hoc analysis showed a significant difference between monaural and binaural sound presentation at low-speed (p < .01) and at high-speed (p < .01) but no significant difference for the no-rotation condition (i.e., static, p = 0.56).

The two-way ANOVA showed significant interaction between the presentation condition and rotation speed for the pleasantness ratings (F(2,36) = 3.86, p < .05, $\eta_p^2$ = 0.18). The main effect of presentation condition (F(1,18) = 4.71, p <.05, $\eta_p^2$ = 0.21) and rotation speed (F(2,36) = 4.11, p < .05, $\eta_p^2$ = 0.19) was significant (Figure 2D). The post-hoc analysis showed a significant difference between the monaural and binaural sound presentation at low-speed (p < .05) but no significant difference for the **Proximal binaural sounds make frisson**no-rotation or and high-speed rotation condition (p = 0.22 and p = 0.069, respectively). Table 1 also presents the result of acoustic feature analysis for music stimuli.

---------------------------------------------- Table 1 about here --------------------------------------------------

---------------------------------------------- Figure 3 about here --------------------------------------------------

---------------------------------------------- Table 2 about here --------------------------------------------------

---------------------------------------------- Table 3 about here --------------------------------------------------

**Discussion**

Our results showed that moving sound stimuli (including both those approaching the participant's head [monaural] and those moving proximally around it [binaural]) can induce magnitudes of subjective frisson that are different from those produced by non-moving sound stimuli. The participants reported a stronger magnitude of frisson when sound stimuli were moving (Figure 2A, Table 2), indicating that subjective ratings for frisson are also velocity dependent. A rotating sound stimulus can be interpreted as a variant of a looming sound as a mean of cross-wise loudness differences in the space. ILDs increased where proximal sounds (within 1 m of the listener's body) approached (Brungart and Rabinowitz, 1999). A prepared rotating sound produced small ILDs (almost none when the speaker was placed in the center between the parts of the binaural microphone) and substantially large ones (when it was placed to one side or the other of the binaural microphone). The recorded sounds could be interpreted to represent lateral sound sources that are approaching and/or receding.

Figure 2B indicates a remarkable decrease in the subjective ratings for pleasantness when sound stimuli rotated. This may indicate that the participants considered sound stimuli to be a form of a warning signal, similar to outcomes reported by a previous study for sound stimuli approaching on a horizontal plane in 3D space (Bach et al., 2009).

Our acoustic feature analysis of non-musical stimuli showed that ILDs may explain subjective magnitudes of frisson. The ILD a sound stimulus was correlated with the subjective magnitude of the frisson produced (Figure 3 and Supplemental Table 3). Although the temporal differences in the pressure levels of the sound in a single ear might also be considered to bear on our results, this possibility can be rejected because the subjective magnitude of frisson was significantly weaker in monaural condition (Figure 2A). ILDs are a representative feature in psychoacoustics (Zwicker and Fastl, 2013), but they have largely been investigated in relation to the perception of spatial sounds. To the best of our knowledge, no study has established a relationship between subjective ratings of frisson and ILDs. Our results indicated a significant correlation between the parameter of loudness and subjective frisson (Supplemental Table 3). The sharpness parameter was also correlated with subjective

PAGE \* 7 \*

**Proximal binaural sounds make frisson**

frisson. Sharpness is one of the acoustic parameters that is known to be inversely associated with pleasantness, therefore the significant negative correlation between sharpness and pleasantness in the binaural condition was not surprising. It was intriguing to observe a significant positive correlation between sharpness and the frisson ratings in both binaural and monaural condition. As described in the introduction, sound frisson occurs when an aversive sound (e.g., the buzz of a mosquito) is approaching, so that sharpness could be another index of the magnitude of frisson ratings. It has been shown that loudness peaks could induce the onset of musical frisson when participants are listening to music (Harrison and Loui, 2014). Another study reported that acoustic features, such as loudness, sharpness, and/or tonality, could explain subjective evaluations of musical frisson (Pamua et al., 2016). The study of sound frisson or even music frisson in terms of ILD, loudness, and sharpness should be pursued further. We should admit that the current correlation analysis was limited to the sound stimuli used in the experiment (nine data points), and the value of acoustic parameters was not uniformly distributed. A set of sound stimuli whose value of acoustic parameters were widely distributed would be needed in a further study.

Movement-induced frisson and pleasantness may be independent of sound categories. We used white noise, sounds made by materials (beads and plastic bag), and music. We found significant differences in the subjective ratings of frisson between static and moving auditory stimuli in our study (Figure 2A and Table 2 show the results for noise stimuli and Figure 2C and Table 3 show the results for musical stimuli). The baseline for subjective pleasantness ratings was higher when the auditory stimulus was music (Figure 2B and D). This may be because noise sounds are intrinsically meaningless and can even be aversive. However, the modulation of motion-induced pleasantness was dependent on the variety of sound (Supplemental Figure 1B). The rotation of the sound stimuli also resulted in a decrease in ratings of pleasantness (Supplemental Figure 1B). Therefore, rotating sound stimuli may increase the magnitude of frisson and decrease the magnitude of pleasantness.

There is a need for further research to investigate whether a moving musical sound stimulus can lead to an increase in subjective frisson. In the present study, we defined sound frisson as the feeling of coldness or shivering in the absence of a physically cold stimulus. On the other hand, as mentioned in the introduction, musical frisson can be closely linked to musical surprise. Both the noise acoustic stimulus including the ASMR-inducing sound and music lead subjective frisson (i.e. Barratt and Davis, 2015; Salimoor et al.,2011), but the experience might be explained by different mechanisms. The difference in behavioral and neural evidence between sound (non-musical) frisson and musical frisson is yet to be fully elucidated. For above the reasons, we analyzed subjective frisson produced by non-musical sound and musical sound, respectively (Figure 2, Supplemental Figure 1 and 2). Here, we also used a single piece of music as an auditory stimulus in our exploratory study. Our results indicated that a piece of music can elicit a stronger frisson while moving than while static (Figure 2C); however, it remained unclear whether it was a music contexture augmented by the addition of a spatial movement or the spatial movement itself increased subjective frisson. In future work, different varieties of music stimuli should be used to answer this question. The effects and correlations of the speed of movement with the induction of subjective frisson also remain to be addressed.

Velocity did not significantly change the magnitude of the frisson or pleasantness so long as the sound stimuli were moving. We speculated that the tingling sensation triggered by ASMR is associated with the perception of frisson. A recent study reported that slow and/or repetitive movements typically produced by the human hand evoked ASMR, or a sensation of tingling or pleasantness in human observers (Barret and Davis, 2015). Our data revealed that the magnitude of frisson can change with increasing velocity of the moving stimulus, based on correlation analysis between loudness (which was modulated by moving velocity) and subjective frisson (Figure 3 and

  

Supplemental Table 3). However, pleasantness was significantly decreased when the sound stimuli were moving (Figure 2B). Therefore, the relationship between ASMR and subjective frisson remains elusive, implying that the spatial motion of the stimulus alone cannot explain the perceptual mechanisms of ASMR.

We also investigated the relationship between impulsive or anxious behavior and subjective frisson. A number of recent studies have found a remarkable link between impulsive feelings and misophonia, which is a dislike or hatred of certain proximal sounds (Schroder, Vulink & Denys, 2013; Edelstein et al., 2013). Although we suspected that subjective frisson relates to emotional states (e.g., impulsiveness and anxiety), we did not find a significant correlation between these factors so far (Supplemental Table 2). The previous study used the Big Five Inventory or the Brief Symptom Inventory for calculating the correlation between the ratings of subjective ASMR intensity (Fredborg et al., 2018) or auditory looming effect (Riskind et al., 2014). These proven personality trait measures should be included in further study.

To summarize, we noted that sound stimuli moving around the proximal area of the head may produce frisson and that the magnitude of the frisson is movement dependent. Induced frisson persists irrespective of sound sources. Additionally, the binaural sound produced a larger magnitude of frisson than the monaural sound, implying that spatiality in sound may be an important factor in sound-based frisson. In future studies, we will address spatial auditory features, such as (variance of) ILD, in music-induced frisson.

**Proximal binaural sounds make frisson**

**Tables**

**Table 1.** Acoustic parameters and means of subjective ratings (frisson/ pleasantness) for each auditory stimulus.

| | Sound | Speed | var(ILD)@128Hz | var(ILD)@1336Hz | Loudness [sone] | | | Sharpness [acum] | | | Tonality [tu] | | | Frisson | Pleasantness |
|---|---|---|---|---|---|---|---|---|---|---|---|---|---|---|---|
| | | | | | Right | Left | average | Right | Left | average | Right | Left | average | | |
| Binaural | Beads | 0 | 5.61 | 0.17 | 19.28 | 19.80 | 19.54 | 1.59 | 1.60 | 1.60 | 0.03 | 0.03 | 0.03 | -0.26 | 0.09 |
| | | Low | 126.32 | 54.07 | 13.39 | 15.56 | 14.48 | 1.72 | 1.78 | 1.75 | 0.03 | 0.04 | 0.04 | 0.80 | -0.45 |
| | | High | 144.72 | 62.29 | 13.82 | 15.07 | 14.45 | 1.71 | 1.77 | 1.74 | 0.03 | 0.04 | 0.04 | 0.88 | -0.59 |
| | Plastic bag | 0 | 0.03 | 0.17 | 31.14 | 32.61 | 31.88 | 1.45 | 1.45 | 1.45 | 0.06 | 0.06 | 0.06 | -0.50 | -0.23 |
| | | Low | 113.44 | 53.47 | 21.74 | 26.14 | 23.94 | 1.58 | 1.67 | 1.63 | 0.06 | 0.06 | 0.06 | 0.24 | -0.43 |
| | | High | 130.86 | 61.77 | 22.43 | 25.28 | 23.86 | 1.62 | 1.70 | 1.66 | 0.06 | 0.05 | 0.06 | 0.78 | -0.37 |
| | White noise | 0 | 0.02 | 0.10 | 33.39 | 34.49 | 33.94 | 1.44 | 1.42 | 1.43 | 0.06 | 0.06 | 0.06 | -0.88 | 0.30 |
| | | Low | 115.92 | 54.03 | 22.85 | 27.56 | 25.21 | 1.57 | 1.65 | 1.61 | 0.06 | 0.06 | 0.06 | 0.27 | -0.12 |
| | | High | 130.43 | 61.68 | 23.87 | 26.77 | 25.32 | 1.59 | 1.66 | 1.63 | 0.06 | 0.06 | 0.06 | 0.42 | -0.18 |
| | Music | 0 | 5.14 | 0.45 | 18.84 | 18.31 | 18.58 | 1.00 | 1.02 | 1.01 | 0.31 | 0.31 | 0.31 | -0.54 | 1.18 |
| | | Low | 117.15 | 53.90 | 16.49 | 16.38 | 16.44 | 1.06 | 1.12 | 1.09 | 0.30 | 0.30 | 0.30 | 0.50 | 0.62 |
| | | High | 131.98 | 61.96 | 15.76 | 16.55 | 16.16 | 1.09 | 1.13 | 1.11 | 0.30 | 0.30 | 0.30 | 0.84 | 0.28 |
| Monaural | Beads | 0 | | | 18.89 | 20.82 | 19.86 | 1.57 | 1.61 | 1.59 | 0.03 | 0.03 | 0.03 | -0.54 | 0.08 |
| | | Low | | | 16.47 | 18.63 | 17.55 | 1.71 | 1.78 | 1.75 | 0.03 | 0.03 | 0.03 | -0.10 | -0.50 |
| | | High | | | 16.39 | 18.63 | 17.51 | 1.71 | 1.77 | 1.74 | 0.02 | 0.03 | 0.03 | 0.30 | -0.34 |
| | Plastic bag | 0 | | | 30.37 | 33.70 | 32.04 | 1.42 | 1.46 | 1.44 | 0.06 | 0.06 | 0.06 | -0.47 | -0.34 |
| | | Low | | | 26.57 | 30.46 | 28.52 | 1.56 | 1.66 | 1.61 | 0.06 | 0.06 | 0.06 | -0.07 | -0.83 |
| | | High | | | 26.52 | 30.52 | 28.52 | 1.60 | 1.70 | 1.65 | 0.05 | 0.05 | 0.05 | -0.19 | -0.63 |
| | White noise | 0 | | | 32.46 | 35.79 | 34.13 | 1.40 | 1.44 | 1.42 | 0.06 | 0.06 | 0.06 | -0.92 | 0.27 |
| | | Low | | | 28.16 | 32.06 | 30.11 | 1.55 | 1.64 | 1.60 | 0.06 | 0.06 | 0.06 | 0.03 | -0.30 |
| | | High | | | 28.29 | 32.29 | 30.29 | 1.57 | 1.66 | 1.62 | 0.06 | 0.06 | 0.06 | 0.17 | -0.31 |
| | Music | 0 | | | 18.28 | 19.31 | 18.80 | 0.98 | 1.01 | 1.00 | 0.31 | 0.30 | 0.31 | -0.38 | 0.92 |
| | | Low | | | 17.99 | 19.24 | 18.62 | 1.06 | 1.11 | 1.09 | 0.31 | 0.30 | 0.31 | -0.26 | 1.16 |
| | | High | | | 17.94 | 19.24 | 18.59 | 1.08 | 1.13 | 1.11 | 0.31 | 0.30 | 0.31 | -0.10 | 0.73 |

**Table 2.** Summary of ANOVA table for non-musical stimuli (frisson/pleasantness ratings). *, **, and *** denote p < 0.05, 0.01, and 0.001, respectively.

| | | | F-value | $\eta_p^2$ | p-value |
|---|---|---|---|---|---|
| Frisson | Interaction | sound type X presentation condition X rotation speed | 1.45 | 0.075 | 0.23 |
| | | sound type X presenation condition | 2.78 | 0.13 | 0.075 |
| | | presentation condition X rotation speed | 2.76 | 0.13 | 0.077 |
| | | rotation speed X sound type | 1.20 | 0.062 | 0.32 |
| | Main effect | sound type | 2.20 | 0.11 | 0.13 |
| | | presentation condition | 10.28 | 0.36 | 0.005 ** |
| | | rotation speed | 43.38 | 0.71 | 0.000 *** |
| Pleasantness | Interaction | sound type X presentation condition X rotation speed | 0.27 | 0.015 | 0.90 |
| | | sound type X presentation condition | 1.38 | 0.071 | 0.26 |
| | | presentation condition X rotation speed | 0.51 | 0.027 | 0.61 |
| | | rotation speed X sound type | 0.55 | 0.029 | 0.70 |
| | Main effect | sound type | 3.45 | 0.16 | 0.043 * |
| | | presentation condition | 2.29 | 0.11 | 0.15 |
| | | rotation speed | 7.73 | 0.30 | 0.002 * |

**Proximal binaural sounds make frisson**

**Table 3.** Summary of ANOVA table for musical stimuli (frisson/pleasantness ratings). *, **, and *** denote p < 0.05, 0.01, and 0.001, respectively.

| | | | F-value | $\eta_p^2$ | p-value |
|---|---|---|---|---|---|
| Frisson | Interaction | presentation condition X rotation speed | 6.39 | 0.26 | 0.004 ** |
| | Main effect | presentation condition | 10.77 | 0.37 | 0.004 ** |
| | | rotation speed | 11.45 | 0.39 | 0.000 *** |
| Pleasantness | Interaction | presentation condition X rotation speed | 3.86 | 0.18 | 0.03 * |
| | Main effect | presentation condition | 4.71 | 0.21 | 0.04 * |
| | | rotation speed | 4.11 | 0.19 | 0.03 * |

**Figure legends**

**Figure 1. Schematic of setup for auditory stimulus recording.** A loudspeaker was placed near to a binaural microphone. The binaural microphone was attached to a stepping motor, which is controlled remotely. The dummy head was fixed (static condition) or rotated at certain velocities while sound sources were played from the loudspeaker.

**Figure 2. Subjective rating scores of auditory stimuli.** (A) The binaural (dichotic) sound induced more frisson than the monaural (diotic) sound did for non-musical (noise) sounds. However, no significant difference was found between monaural and binaural sounds in the pleasantness rating (B). The factors that were found to be significantly different in the three-way repeated ANOVA are indicated as follows: *p < .05, **p < .01, and *** p < .001. The white and black bars represent the results for the monaural and binaural sounds, respectively. We used three rotational conditions: no rotation [static] (0°/s), low speed (18°/s), and high speed (72°/s).
(C) and (D) show the subjective rating scores for the music. The binaural sounds induced a higher subjective frisson rating than the monaural sound did under rolling conditions (C). However, the binaural sound decreased the subjective rating of pleasantness more than the monaural sound did (D). Significantly different factors, determined through two-way repeated ANOVA, were indicated. The error bars indicate standard error of the mean.

**Figure 3. The relationship between subjective ratings (frisson/pleasantness) and acoustic features.** We conducted a correlation analysis only for non-musical (noise) stimuli (see also Supplemental Table 3). The ILD variance (calculated only for binaural sounds) for the 128 and 1338 Hz bands of sound stimuli was correlated with the subjective magnitude of the frisson (A, B); moreover, we saw a significant correlation between subjective frisson and loudness for binaural sound stimuli (C). Similarly, the variance of ILD in the 128 Hz and 1338 Hz bands of sound stimuli was correlated with the subjective magnitude of pleasantness (D, E), and we also found a significant correlation between pleasantness and loudness only for binaural sound stimuli (F). The dashed line on the scatter plot represents the line of best fit regression line if there was a significant relationship between acoustic features and subjective ratings. *r*, *p*, and *n.s.* denote the Pearson's product moment correlation coefficient, p-value, and not significant, respectively.

**Proximal binaural sounds make frisson**

**Supplemental Figure legends**

**Supplemental Figure 1**
Subjective ratings of noise stimuli (rolling beads, rubbing a plastic bag, white noise). The main effect of the sound type was not significant in the frisson ratings ($F(2,36) = 2.20$, $p = 0.13$, $\eta_p^2 = 0.11$), but significant in the pleasantness ratings ($F(2,36) = 3.45$, $p < .05$, $\eta_p^2 = 0.16$). The post-hoc analysis showed a significant difference between the plastic bag and white noise sounds. * and n.s. denotes $p < .05$ and not significant. Error bar indicates standard error.

**Supplemental Figure 2**
A comparison between different sound pressure level conditions for noise stimuli (A and B) and for musical stimuli (C and D). Comparing with the normal sound pressure level condition (Normal), the level was decreased -30 dB in the soft condition (Soft) and was increased +2 dB in the loud condition (Loud). We conducted a supplemental experiment comparing these conditions in different groups of participants (N=6 in soft, N=6 in loud conditions). Mixed design repeated ANOVA (three sound pressure levels (soft, normal, and loud) × three rotation speeds (static (no rotation), low-speed and high-speed) showed there was not significant effect of sound pressure levels either on noise or musical stimuli on subjective frisson ($F(2, 90) = 0.71$, $p = 0.50$, $\eta_p^2 = 0.02$ and $F(2, 28) = 1.99$, $p = 0.16$, $\eta_p^2 = 0.13$) and in pleasantness ratings ($F(2, 90) = 0.48$, $p = 0.62$, $\eta_p^2 = 0.01$ and $F(2, 28) = 1.40$, $p = 0.25$, $\eta_p^2 = 0.09$). We used the normal sound pressure level condition in the main experiment.

**Supplemental Figure 3**
Distributions of A-weighted sound pressure levels (dBA) in both left and right channels for each auditory stimulus. Red lines, thick and thin black bars indicate mean, standard error and deviation, and white circles indicate the data points from twenty four auditory stimuli (Supplementary Table 1). The 5th, 50th, and 95th percentile values of equivalent sound pressure levels are shown for the left and right ears (labeled as LAeq). We also show the maximum sound levels computed with the "slow" and "fast" time weighting (labeled as LAS and LAF respectively) sampled every second from each auditory stimulus.

**Proximal binaural sounds make frisson**

**Supplemental Table legends**

**Supplemental Table 1**
Distributions of A-weighted sound pressure levels sampled every second for each stimulus. The 5th, 50th, and 95th percentile values of equivalent sound pressure levels are shown for the left and right ears (labeled as LAeq). We also show the maximum levels Equivalent continuous A-weighted sound pressure level (LAsq) for each auditory stimulus used in the experiment. We computed with the "slow" and "fast" time weighting (labeled as LASmax and LAFmax respectively). We confirmed that all the auditory stimuli were confirmed that all the auditory stimuli were below 85 dB, which is allowed noise level for 8 hours exposure (US Department of Health and Human below 85 dB for LAeq and LASmax, which is allowed noise level for 8 hours exposure (US Department of Health and Human Services. 1998). For Services. 1998). LAFmax, three auditory stimuli exceeded 85 dB but these were less than 90 dB, which is allowed noise level for 2.5 hours exposure.

**Supplemental Table 2**
Results of correlation analysis between subjective frisson rating and psychological assessment (State-Trait Anxiety Inventory-JYZ: STAI and UPPS-P Impulsive Behavior Inventory in Japanese). A significant correlation was not found between these parameters.

**Supplemental Table 3**
Results of correlation analysis between subjective rating (frisson / pleasantness) and acoustic parameters (Figure 3 shows scatter plots of these results) in the condition of non-musical sound (noise) stimuli. There were significant correlations between the variance of ILDs at specific frequency and the magnitude of subjective frisson rating. Loudness and sharpness were also significantly correlated with subjective frisson and pleasantness rating in several conditions (*, **, and *** denote $p < 0.05$, 0.01, and 0.001, respectively).

**Proximal binaural sounds make frisson**

**Supplemental Material**

*Supplemental figures and tables are provided as follows:*

Image1.TIF: Supplemental Figure 1

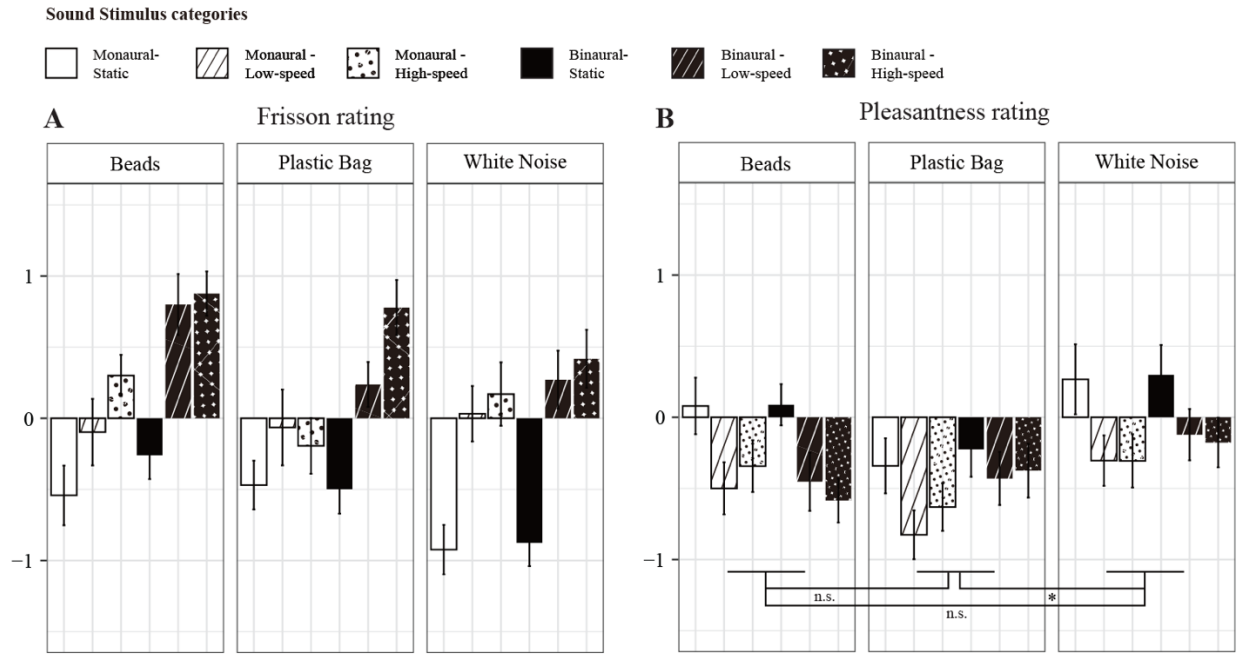

**Supplemental Figure 1** | Subjective ratings of noise stimuli (rolling beads, rubbing a plastic bag, white noise). The main effect of the sound type was not significant in the frisson ratings ($F(2,36) = 2.20$, $p = 0.13$, $\eta_p^2 = 0.11$), but significant in the pleasantness ratings ($F(2,36) = 3.45$, $p < .05$, $\eta_p^2 = 0.16$). The post-hoc analysis showed a significant difference between the plastic bag and white noise sounds. * and n.s. denotes $p < .05$ and not significant. Error bar indicates standard error.

**Proximal binaural sounds make frisson**

Image2.TIF: Supplemental Figure 2

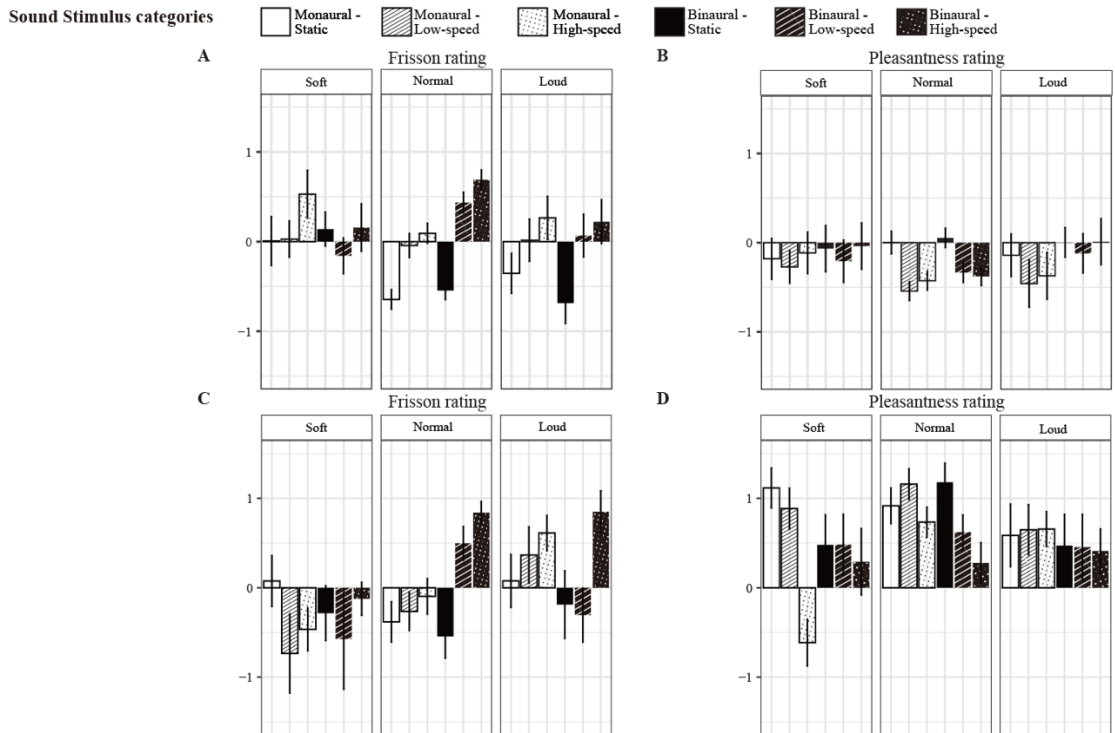

**Supplemental Figure 2** | A comparison between different sound pressure level conditions for noise stimuli (A and B) and for musical stimuli (C and D). Comparing with the normal sound pressure level condition (Normal), the level was decreased -30 dB in the soft condition (Soft) and was increased +2 dB in the loud condition (Loud). We conducted a supplemental experiment comparing these conditions in different groups of participants (N=6 in soft, N=6 in loud conditions). Mixed design repeated ANOVA (three sound pressure levels (soft, normal, and loud) × three rotation speeds (static (no rotation), low-speed and high-speed) showed there was not significant effect of sound pressure levels either on noise or musical stimuli on subjective frisson ($F(2, 90) = 0.71$, $p = 0.50$, $\eta_p^2 = 0.02$ and $F(2, 28) = 1.99$, $p = 0.16$, $\eta_p^2 = 0.13$) and in pleasantness ratings ($F(2, 90) = 0.48$, $p = 0.62$, $\eta_p^2 = 0.01$ and $F(2, 28) = 1.40$, $p = 0.25$, $\eta_p^2 = 0.09$). We used the normal sound pressure level condition in the main experiment.

**Proximal binaural sounds make frisson**

Image3.TIF: Supplemental Figure 3

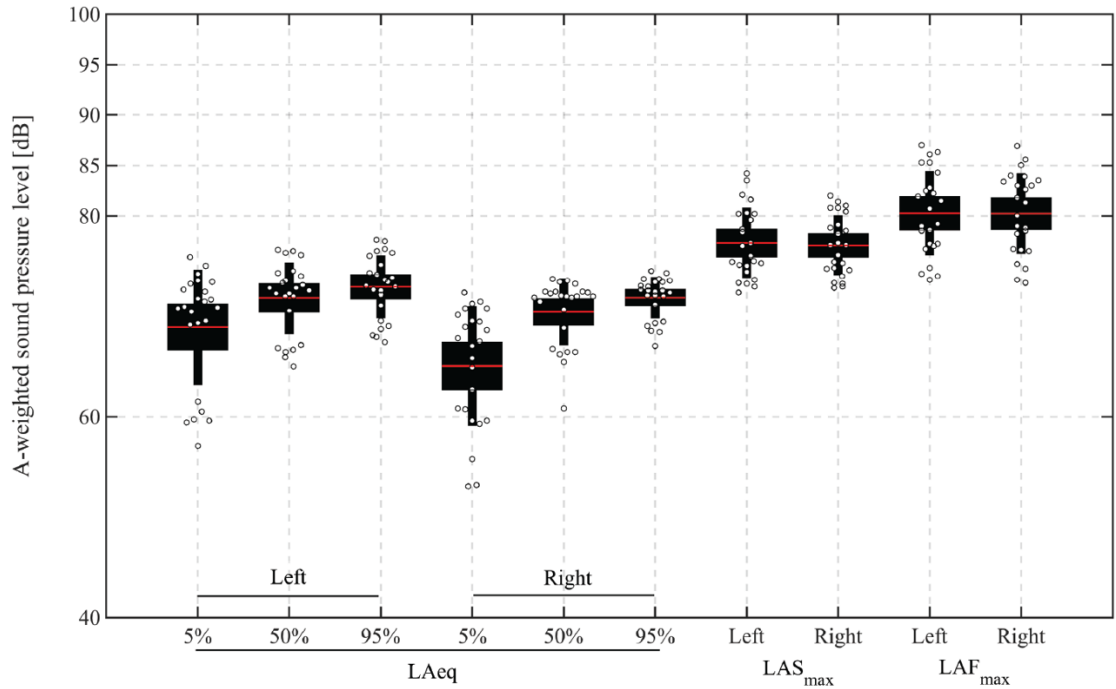

**Supplemental Figure 3** | Distributions of A-weighted sound pressure levels (dBA) in both left and right channels for each auditory stimulus. Red lines, thick and thin black bars indicate mean, standard error and deviation, and white circles indicate the data points from twenty four auditory stimuli (Supplementary Table 1). The 5th, 50th, and 95th percentile values of equivalent sound pressure levels are shown for the left and right ears (labeled as LAeq). We also show the maximum sound levels computed with the "slow" and "fast" time weighting (labeled as LAS and LAF respectively) sampled every second from each auditory stimulus.

# Proximal binaural sounds make frisson

Image4.TIF: Supplemental Table 1

**SUPPLEMENTAL TABLE 1** | Distributions of A-weighted sound pressure levels sampled every second for each stimulus. The 5th, 50th, and 95th percentile values of equivalent sound pressure levels are shown for the left and right ears (labeled as LAeq). We also show the maximum levels computed with the "slow" and "fast" time weighting (labeled as LASmax and LAFmax respectively). We confirmed that all the auditory stimuli were below 85 dB for LAeq and LASmax, which is allowed noise level for 8 hours exposure (US Department of Health and Human Services. 1998). For LAFmax, three auditory stimuli exceeded 85 dB but these were less than 90 dB, which is allowed noise level for 2.5 hours exposure.

| | | | LAeq | | | | | | LASmax | | LAFmax | |
|---|---|---|---|---|---|---|---|---|---|---|---|---|
| | Sound | Speed | Left | | | Right | | | Left | Right | Left | Right |
| | | | 5 | 50 | 95 | 5 | 50 | 95 | | | | |
| Binaural | Beads | 0 | 70.7 | 71.9 | 72.1 | 67.8 | 71.8 | 72.0 | 75.4 | 75.4 | 79.0 | 79.0 |
| | | Low | 70.9 | 71.9 | 74.3 | 60.8 | 71.4 | 73.0 | 78.4 | 82.0 | 81.9 | 86.9 |
| | | High | 69.1 | 70.5 | 71.0 | 60.7 | 68.8 | 71.1 | 77.0 | 79.1 | 82.5 | 85.0 |
| | Plastic bag | 0 | 70.4 | 72.8 | 73.0 | 70.1 | 72.4 | 72.6 | 75.0 | 74.8 | 77.0 | 76.7 |
| | | Low | 72.6 | 74.3 | 76.0 | 62.7 | 72.2 | 73.7 | 80.2 | 81.4 | 82.2 | 83.4 |
| | | High | 73.2 | 73.5 | 74.1 | 70.7 | 72.4 | 72.9 | 78.7 | 78.8 | 82.8 | 84.0 |
| | White noise | 0 | 69.3 | 72.2 | 72.6 | 65.8 | 72.0 | 72.5 | 73.2 | 73.2 | 73.6 | 73.6 |
| | | Low | 73.5 | 74.5 | 76.5 | 64.9 | 71.9 | 73.4 | 80.3 | 80.8 | 81.5 | 83.0 |
| | | High | 72.4 | 73.3 | 73.6 | 68.9 | 71.8 | 72.4 | 77.3 | 77.3 | 80.7 | 81.8 |
| | Music | 0 | 59.4 | 65.9 | 68.1 | 59.6 | 66.7 | 69.0 | 72.3 | 73.3 | 74.2 | 75.2 |
| | | Low | 59.7 | 66.8 | 67.9 | 53.1 | 60.8 | 69.3 | 73.3 | 77.1 | 76.7 | 80.0 |
| | | High | 60.5 | 65.0 | 67.4 | 53.2 | 65.4 | 67.0 | 75.1 | 74.7 | 78.6 | 78.2 |
| Monaural | Beads | 0 | 71.7 | 72.2 | 72.5 | 67.0 | 71.9 | 72.0 | 75.5 | 75.3 | 79.2 | 78.8 |
| | | Low | 71.4 | 74.0 | 75.1 | 69.5 | 72.4 | 73.4 | 82.1 | 80.8 | 87.0 | 85.6 |
| | | High | 71.6 | 72.8 | 73.8 | 67.5 | 70.6 | 71.9 | 79.6 | 78.2 | 85.3 | 83.9 |
| | Plastic bag | 0 | 70.8 | 73.1 | 73.4 | 69.4 | 72.3 | 72.5 | 75.3 | 74.6 | 77.2 | 76.5 |
| | | Low | 73.4 | 76.6 | 77.6 | 70.7 | 73.7 | 74.5 | 84.2 | 81.0 | 86.1 | 83.0 |
| | | High | 75.9 | 76.3 | 76.7 | 72.3 | 73.4 | 73.7 | 81.6 | 78.5 | 86.3 | 83.5 |
| | White noise | 0 | 69.5 | 72.5 | 72.9 | 68.6 | 71.9 | 72.3 | 73.6 | 72.9 | 74.0 | 73.3 |
| | | Low | 74.2 | 76.5 | 77.5 | 71.2 | 73.5 | 74.3 | 83.5 | 80.4 | 85.3 | 82.6 |
| | | High | 75.0 | 76.1 | 76.3 | 71.4 | 73.2 | 73.5 | 80.2 | 77.1 | 84.3 | 81.3 |
| | Music | 0 | 59.6 | 66.4 | 68.7 | 59.3 | 66.2 | 68.5 | 72.9 | 72.8 | 74.8 | 74.7 |
| | | Low | 57.1 | 66.6 | 69.5 | 55.8 | 66.4 | 69.4 | 76.0 | 76.1 | 78.5 | 78.5 |
| | | High | 61.5 | 67.1 | 69.0 | 59.6 | 66.4 | 68.4 | 74.4 | 74.0 | 77.2 | 76.6 |

Image5.TIF: Supplemental Table 2

**SUPPLEMENTAL TABLE 2** | Results of correlation analysis between subjective frisson rating and psychological assessment (State-Trait Anxiety Inventory-JYZ: STAI and UPPS-P Impulsive Behavior Inventory in Japanese). A significant correlation was not found between these parameters.

| Personality trait measures | Music stimuli | | Noise Stimuli | |
| --- | --- | --- | --- | --- |
| | coefficient | p value | coefficient | p value |
| Premeditation (lack of) | 0.09 | 0.7 | 0.24 | 0.31 |
| Urgency | 0.03 | 0.9 | -0.08 | 0.74 |
| Sensation Seeking | 0.10 | 0.67 | 0.03 | 0.89 |
| Perservance (lack of) | 0.32 | 0.18 | -0.15 | 0.55 |
| STAI (State) | 0.13 | 0.59 | -0.35 | 0.14 |
| STAI (Trait) | 0.19 | 0.44 | -0.28 | 0.25 |

Image6.TIF: Supplemental Table 3

**SUPPLEMENTAL TABLE 3** | Results of correlation analysis between subjective rating (frisson / pleasantness) and acoustic parameters (Figure 3 shows scatter plots of these results) in the condition of non-musical sound (noise) stimuli. There were significant correlations between the variance of ILDs at specific frequency and the magnitude of subjective frisson rating. Loudness and sharpness were also significantly correlated with subjective frisson and pleasantness rating in several conditions (*, **, and *** denote $p < 0.05$, 0.01, and 0.001, respectively).

| Sensory evaluation [sound types] | | ILD(128Hz) | | ILD(1336Hz) | | Loudness | | Sharpness | | Tonality | |
| --- | --- | --- | --- | --- | --- | --- | --- | --- | --- | --- | --- |
| | | coefficient | p value | coefficient | p value | coefficient | p value | coefficient | p value | coefficient | p value |
| Frisson | [monaural] | — | — | — | — | -0.39 | 0.30 | 0.77 | 0.015 * | -0.22 | 0.58 |
| | [binaural] | 0.93 | 0.000 *** | 0.91 | 0.001 ** | -0.78 | 0.013 * | 0.95 | 0.000 *** | -0.33 | 0.38 |
| Pleasantness | [monaural] | — | — | — | — | 0.12 | 0.77 | -0.52 | 0.15 | -0.07 | 0.85 |
| | [binaural] | -0.76 | 0.018 * | -0.74 | 0.024 * | 0.68 | 0.065 | -0.76 | 0.019 * | 0.22 | 0.57 |

*Examples of sound stimuli used in this study:*

Audio1.wav: Binaural sound of static microphones

Audio2.wav: Monaural sound of moving microphones (72°/s)

Audio3.wav: Binaural sound of moving microphones (72°/s)

You can hear these sounds at sound cloud:

https://soundcloud.com/sfctouchlab/sets/sound-samples-for-proximal-1

**Proximal binaural sounds make frisson**

## Author Contributions

SH and YI conducted experiments, data analysis and manuscript writing. RK, EM, NN, KS conducted recording procedures including experiment execution. TK, HK, SF conceived evaluation experiments. All authors conceived and wrote this manuscript and SF and MN supervised this research project.

## References

Altman, J.A., and Romanov, V.P. (1988). Psychophysical characteristics of the auditory image movement perception during dichotic stimulation. *Int J Neurosci* 38(3-4), 369-379. doi: 10.3109/00207458808990697.

Bach, D.R., Neuhoff, J.G., Perrig, W., and Seifritz, E. (2009). Looming sounds as warning signals: the function of motion cues. *Int J Psychophysiol* 74(1), 28-33. doi: 10.1016/j.ijpsycho.2009.06.004.

Barratt, E.L., and Davis, N.J. (2015). Autonomous Sensory Meridian Response (ASMR): a flow-like mental state. *PeerJ* 3, e851. doi: 10.7717/peerj.851.

Barratt, E.L., Spence, C., and Davis, N.J. (2017). Sensory determinants of the autonomous sensory meridian response (ASMR): understanding the triggers. *PeerJ* 5, e3846. doi: 10.7717/peerj.3846.

Blood, A.J., and Zatorre, R.J. (2001). Intensely pleasurable responses to music correlate with activity in brain regions implicated in reward and emotion. *Proc Natl Acad Sci U S A* 98(20), 11818-11823. doi: 10.1073/pnas.191355898.

Brungart, D.S., and Rabinowitz, W.M. (1999). Auditory localization of nearby sources. Head-related transfer functions. *J Acoust Soc Am* 106(3 Pt 1), 1465-1479.

Carlile, S., and Leung, J. (2016). The Perception of Auditory Motion. *Trends Hear* 20. doi: 10.1177/2331216516644254.

Edelstein, M., Brang, D., Rouw, R., and Ramachandran, V.S. (2013). Misophonia: physiological investigations and case descriptions. *Front Hum Neurosci* 7, 296. doi: 10.3389/fnhum.2013.00296.

Fredborg, B., Clark, J., and Smith, S.D. (2017). An Examination of Personality Traits Associated with Autonomous Sensory Meridian Response (ASMR). *Front Psychol* 8, 247. doi: 10.3389/fpsyg.2017.00247.

Harrison, L., and Loui, P. (2014). Thrills, chills, frissons, and skin orgasms: toward an integrative model of transcendent psychophysiological experiences in music. *Front Psychol* 5, 790. doi: 10.3389/fpsyg.2014.00790.

Huron, D., and Hellmuth, E. (2011). Musical expectancy and thrills.

**Proximal binaural sounds make frisson**

Kondo, H.M., Farkas, D., Denham, S.L., Asai, T., and Winkler, I. (2017). Auditory multistability and neurotransmitter concentrations in the human brain. *Philos Trans R Soc Lond B Biol Sci* 372(1714). doi: 10.1098/rstb.2016.0110.

McDermott, J.H., Schemitsch, M., and Simoncelli, E.P. (2013). Summary statistics in auditory perception. *Nat Neurosci* 16(4)**,** 493-498. doi: 10.1038/nn.3347.

McCarthy, L., Olsen K.N. (2017) A "looming bias" in spatial hearing? Effects of acoustic intensity and spectrum on categorical sound source localization. *Atten Percept Psychophys* 79(1), 352-362. doi: 10.3758/s13414-016-1201-9.

Neuhoff, J.G. (2016). Looming sounds are perceived as faster than receding sounds. *Cogn Res Princ Implic* 1(1)**,** 15. doi: 10.1186/s41235-016-0017-4.

Okada,S, Hirahara,T.(2015).Perception of approaching and retreating sounds. *Acoust Sci Technol* 36(5), 449-452. doi: https://doi.org/10.1250/ast.36.449

Pamua, H., Kasprzak, C., and Kłaczyński, M. (2016). Nuisance assessment of different annoying sounds based on psychoacoustic metrics and electroencephalography. *Diagnostyka* 17(2)**,** 8.

Riskind, J.H., Kleiman, E.M., Seifritz, E., and Neuhoff, J. (2014). Influence of anxiety, depression and looming cognitive style on auditory looming perception. *J Anxiety Disord* 28(1)**,** 45-50. doi: 10.1016/j.janxdis.2013.11.005.

Salimpoor, V.N., Benovoy, M., Larcher, K., Dagher, A., and Zatorre, R.J. (2011). Anatomically distinct dopamine release during anticipation and experience of peak emotion to music. *Nat Neurosci* 14(2)**,** 257-262. doi: 10.1038/nn.2726.

Schroder, A., Vulink, N., and Denys, D. (2013). Misophonia: diagnostic criteria for a new psychiatric disorder. *PLoS One* 8(1)**,** e54706. doi: 10.1371/journal.pone.0054706.

Sloboda, J.A. (1991). Music structure and emotional response: Some empirical findings. *Psychology of music* 19(2)**,** 110-120.

Spielberger, C.D. (2010). State-Trait anxiety inventory. *The Corsini encyclopedia of psychology***,** 1-1.

Tajadura-Jimenez, A., Valjamae, A., Asutay, E., and Vastfjall, D. (2010). Embodied auditory perception: the emotional impact of approaching and receding sound sources. *Emotion* 10(2)**,** 216-229. doi: 10.1037/a0018422.

US Department of Health and Human Services. (1998). Criteria for a recommended standard: occupational noise exposure. Revised criteria 1998. Centers for Disease Control and Prevention, National Institute for Occupational Safety and Health.

Whiteside, S.P., and Lynam, D.R. (2001). The five factor model and impulsivity: Using a structural model of personality to understand impulsivity. *Personality and individual differences* 30(4)**,** 669-689.

**Proximal binaural sounds make frisson**

Zwicker, E., and Fastl, H. (2013). *Psychoacoustics: Facts and models.* Springer Science & Business Media.

**Sound Stimulus categories**

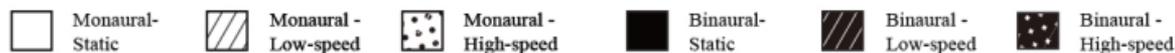

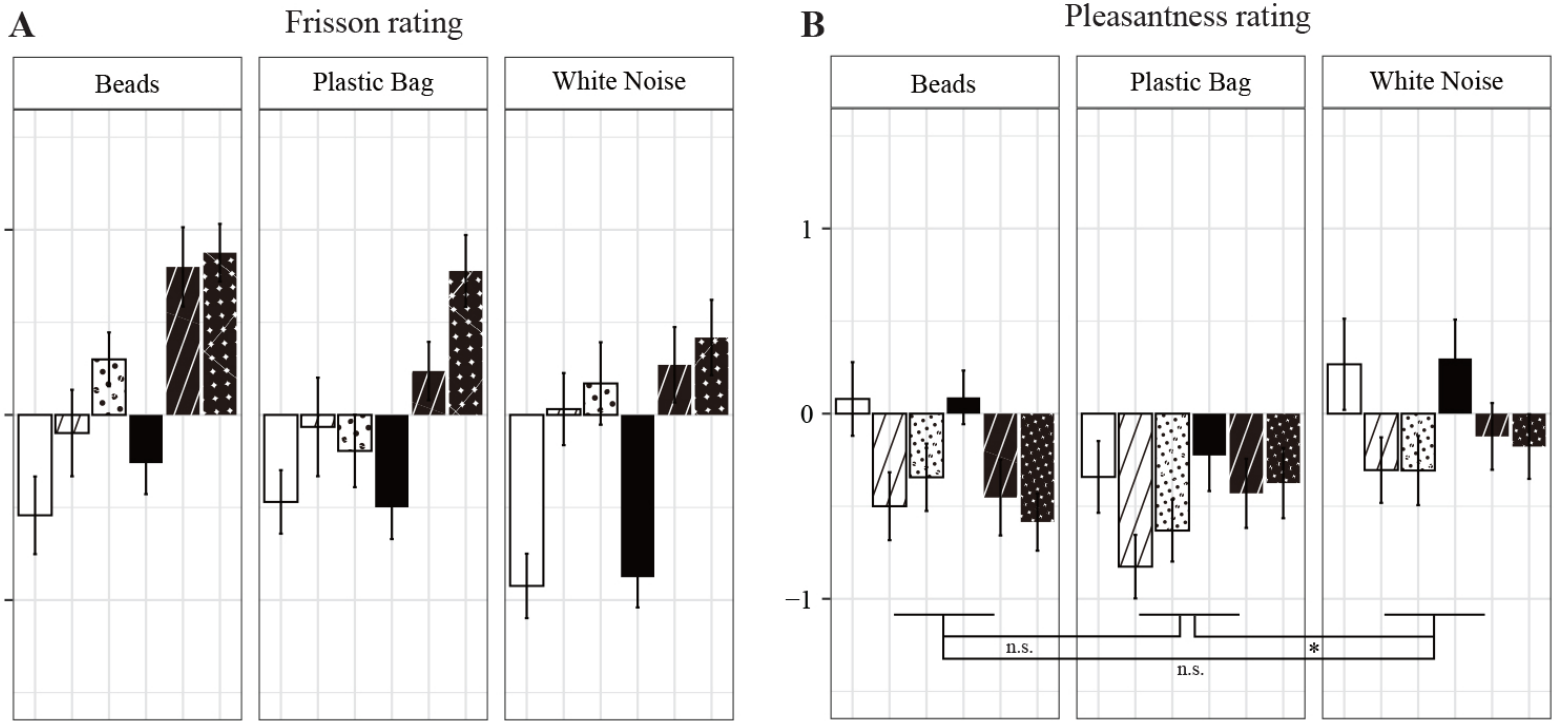

**Supplemental Figure 1** | Subjective ratings of noise stimuli (rolling beads, rubbing a plastic bag, white noise). The main effect of the sound type was not significant in the frisson ratings ($F(2,36) = 2.20$, $p = 0.13$, $\eta_p^2 = 0.11$), but significant in the pleasantness ratings ($F(2,36) = 3.45$, $p < .05$, $\eta_p^2 = 0.16$). The post-hoc analysis showed a significant difference between the plastic bag and white noise sounds. * and n.s. denotes $p < .05$ and not significant. Error bar indicates standard error.

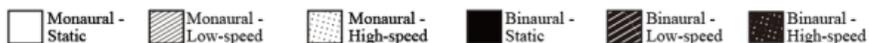
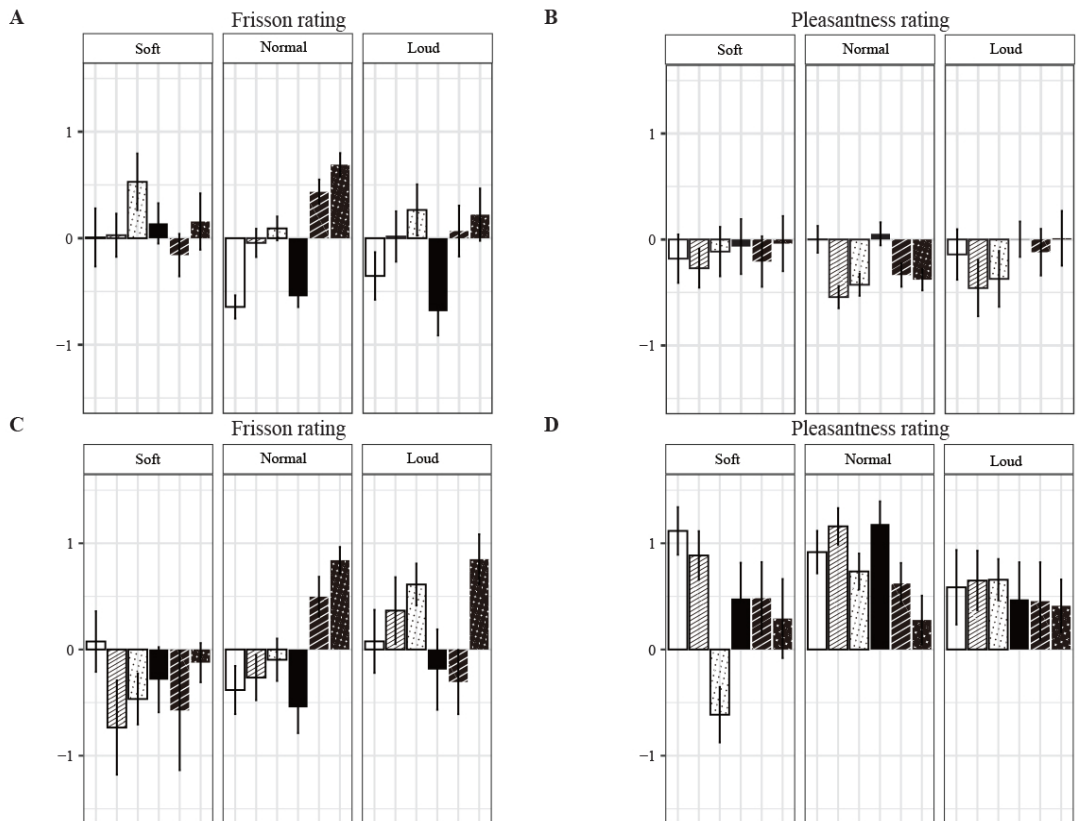

**Supplemental Figure 2** | A comparison between different sound pressure level conditions for noise stimuli (A and B) and for musical stimuli (C and D). Comparing with the normal sound pressure level condition (Normal), the level was decreased -30 dB in the soft condition (Soft) and was increased +2 dB in the loud condition (Loud). We conducted a supplemental experiment comparing these conditions in different groups of participants (N=6 in soft, N=6 in loud conditions). Mixed design repeated ANOVA (three sound pressure levels (soft, normal, and loud) × three rotation speeds (static (no rotation), low-speed and high-speed) showed there was not significant effect of sound pressure levels either on noise or musical stimuli on subjective frisson ($F(2, 90) = 0.71$, $p = 0.50$, $\eta_p^2 = 0.02$ and $F(2, 28) = 1.99$, $p = 0.16$, $\eta_p^2 = 0.13$) and in pleasantness ratings ($F(2, 90) = 0.48$, $p = 0.62$, $\eta_p^2 = 0.01$ and $F(2, 28) = 1.40$, $p = 0.25$, $\eta_p^2 = 0.09$). We used the normal sound pressure level condition in the main experiment.

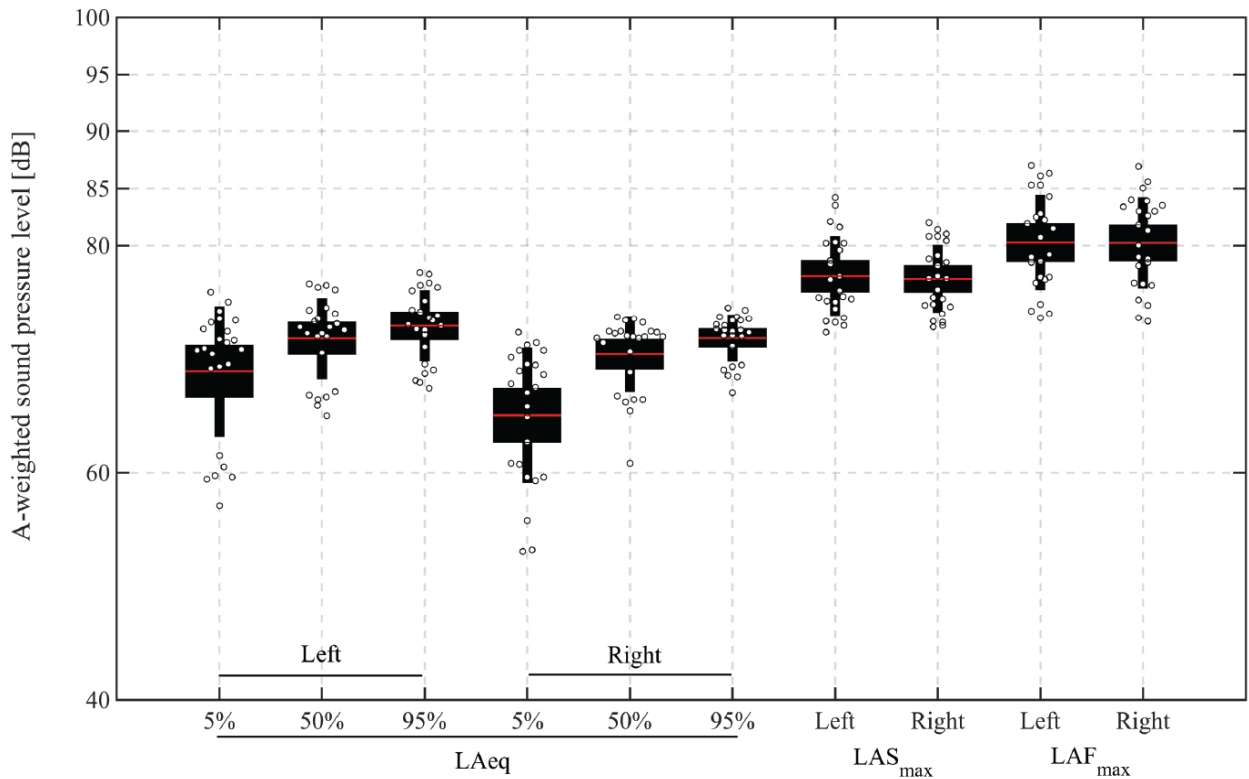

**Supplemental Figure 3** | Distributions of A-weighted sound pressure levels (dBA) in both left and right channels for each auditory stimulus. Red lines, thick and thin black bars indicate mean, standard error and deviation, and white circles indicate the data points from twenty four auditory stimuli (Supplementary Table 1). The 5th, 50th, and 95th percentile values of equivalent sound pressure levels are shown for the left and right ears (labeled as LAeq). We also show the maximum sound levels computed with the "slow" and "fast" time weighting (labeled as LAS and LAF respectively) sampled every second from each auditory stimulus.